\documentstyle[psfig,aps,twocolumn]{revtex}

\tightenlines

\def\beq{\begin{equation}}
\def\eeq{\end{equation}}
\def\bea{\begin{eqnarray}}
\def\eea{\end{eqnarray}}

\begin{document}
\draft

\title{Absence of Thermophoretic Flow in Relativistic Heavy-Ion Collisions as 
an Indicator for the Absence of a Mixed Phase}

\author{Markus H. Thoma}
\address{Max-Planck-Institut f\"ur extraterrestrische Physik, 
Giessenbachstra{\ss}e, 85748 Garching, Germany}

\date{\today}

\maketitle

\begin{abstract}
If a quark-gluon plasma is formed in relativistic heavy-ion collisions,
there may or may not be a mixed phase of quarks, gluons and hadronic
clusters when the critical temperature is reached in the expansion of
the fireball. If there is a temperature gradient in the fireball, the
hadronic clusters, embedded in the heat bath of quarks and gluons,
would be subjected to a thermophoretic force. It is shown that, even
for small temperature gradients and short lifetimes of the mixed phase,
thermophoresis would lead to a flow essentially stronger than the
observed one. The absence of this strong flow provides 
support for a rapid or sudden hadronization mechanism without a mixed
phase.

\end{abstract} 

\bigskip

%{\hspace*{1.1cm}Keywords: Energy loss; Relativistic plasmas; Thermal Field 
%Theory}
%\medskip

\narrowtext
%\newpage

%\section{Introduction}

Relativistic heavy-ion collision experiments at CERN (SPS, LHC) and BNL (RHIC) 
are dedicated to the discovery of the quark-gluon plasma, a hypothetic state of matter
in which quarks and gluons are deconfined. The basic idea is to create a hot 
and dense fireball in a high-energy nucleus-nucleus collision with a temperature
above the critical one of the order $T_c=170$ MeV, predicted by QCD lattice computations
\cite{Karsch01}. This tiny fireball will rapidly 
expand and reach the transition temperature after a few fm/c. Depending on the order 
of the deconfinement phase transition there might be a mixed phase of quarks, 
gluons and hadrons. QCD lattice simulations of the deconfinement phase transition
show either a continuous (cross over) or a first order phase transition depending
crucially on the masses of the dynamical quarks \cite{Karsch01}. 

Although the question of the order of the phase transition is still open\cite{Karsch01},
let us assume that there is a first order phase transition from the quark-gluon 
plasma to the hadronic matter leading to a mixed phase. Possible consequences of
a mixed phase in relativistic heavy-ion collisions have been extensively studied
(see e.g. Ref.\cite{Rischke96}). Here we want to discuss a new aspect of the 
mixed phase, namely the possibility of a thermophoretic flow. 
 
Since hadronization does not take place instantaneously -- although
hadronziation may proceed rapidly \cite{Csernai95} -- 
the hadronic bubbles might contain confined clusters of quarks instead of
the final hadrons. These clusters or pre-hadrons as well as usual hadrons
are heavy and extended objects embedded in the thermal quark-gluon plasma phase 
similar as dust particles in a low-temperature plasma \cite{Bouchoule99}.   

It is reasonable to assume that the temperature of the fireball is higher at the 
center than at the surface of the fireball. For example, in hydrodynamical 
calculations a local maximum temperature differing from the average temperature by
as much as 70 MeV has been taken to explain photon data at SPS \cite{Huovinen01}.
Hence, we presuppose the existence of a temperature gradient in the fireball.
For simplicity, we assume a constant temperature gradient from the center
to the surface of a spherical fireball. Owing to this temperature gradient 
in the thermal heat bath of the quark-gluon plasma a thermophoretic force will act 
on the hadronic clusters similar as on dust particles in a low-temperature plasma
\cite{Rothermel01}. This force will push the hadrons from the center to the 
surface resulting in a radial outward motion of hadrons as sketched in Fig.1. Here we assumed 
furthermore that the hadronic bubbles are not transparent for the thermal 
partons but that elastic scattering of the partons off the pre-hadrons is dominant. 

\begin{figure}
\centerline{\psfig{figure=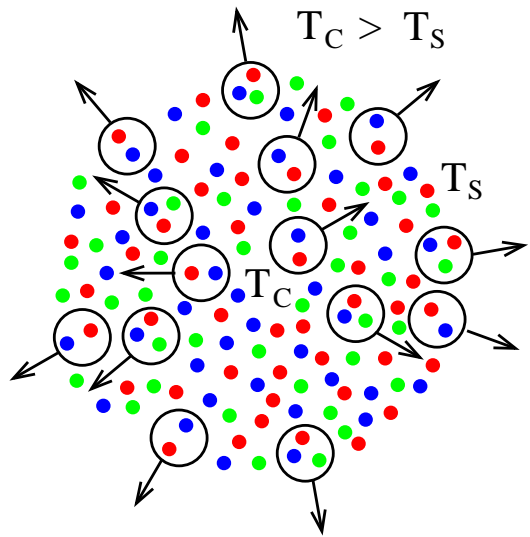,width=7cm}}
\vspace*{0.5cm}
\caption{Sketch of thermophoresis in the mixed phase of the fireball in 
relativistic heavy-ion collisions acting on hadrons. The small dots are the partons
which are the thermal background for the hadrons or quark clusters indicated by the 
circles.}
\end{figure}

A temperature gradient in the mixed phase, for which the temperature is given by the critical
temperature, could be realized in two different ways. First, a finite chemical potential
increasing from the center to the surface would imply a decreasing critical temperature.
Second, if there is a constant chemical potential, a temperature gradient in the fireball 
leads to a mixed-phase shell moving outward. Due to the finite mean free path the partons
from the hotter inner regions can interact with the hadrons in the mixed-phase shell. These 
possibilities could be considered using hydrodynamical methods \cite{Huovinen02}.

The collective motion of hadrons, called flow, is an important feature of heavy-ion collisions
which has been investigated theoretically and experimentally in great detail, providing
important informations on the equation of state of the fireball and especially
a possible signature for the quark-gluon plasma formation \cite{Ollitrault98}.
The different flow patterns (e.g. elliptic and radial flow), studied so far, 
are caused by a pressure gradient in the fireball. Here we propose a new source for
flow, namely a temperature gradient in the fireball. This directed, collective motion of 
hadrons superimposes the usual hydrodynamical flow. 

Using elementary kinetic arguments, the thermophoretic force can be estimated 
similar as the heat conductivity \cite{Reif65}. The latter
follows from the energy flux in the presence of a temperature gradient.
Instead of the energy transfer we consider the momentum transfer per unit time 
and area in a relativistic plasma. Then the thermophoretic force $F$ per area 
$A$ is given as 
\begin{equation}
\frac{F}{A}=-\frac{1}{3}\> n\> \langle u\rangle \> \lambda \>
\frac{d\langle p\rangle}{dT}\> \frac{dT}{dr},
\end{equation}
where $n$ is the number density of the quark-gluon plasma, $\langle u\rangle \simeq 1$ 
the thermal velocity of the partons, $\lambda $ their mean free path, and 
$\langle p\rangle \simeq 3T$ their average momentum.

Hence, the thermophoretic force acting on a hadron is given by
\begin{equation}
F=-\pi \> n\> \lambda \> R^2\> \frac{dT}{dr},
\end{equation}
where $R$ is the hadron radius. Using typical values for the mean free path and the density
of the partons at $T=T_c$, $\lambda =1$ fm, 
$n=5$ fm$^{-3}$, and for the hadron radius, $R=1$ fm, and 
assuming, for example, a small temperature gradient
of $dT/dr=1$ MeV/fm, we get $F=-16$ MeV/fm.

Now, we want to calculate the final velocity $v_f$ of a hadron with mass $m$.
Using
\begin{equation}
-F=\frac{d}{dt}\> \frac{mv}{\sqrt{1-v^2}}=\frac{d}{dx}\> \frac{mv^2}{\sqrt{1-v^2}}
\end{equation}
we find
\begin{equation}
W=-F\> d=\frac{mv_f^2}{\sqrt{1-v_f^2}}
\end{equation}
with the fireball radius $d$. Solving this equation for $v_f$ yields
\begin{equation}
v_f=\frac{W}{\sqrt{2}m}\> \left (\sqrt{1+\frac{4m^2}{W^2}}-1\right )^{1/2}.
\end{equation}

For pions ($m=140$ MeV) and $d=10$ fm we obtain $v_f=0.81$ and for protons 
($m=940$ MeV) $v_f=0.40$, i.e. a strong flow larger or at least comparable
to the hydrodynamic flow as extracted from two-particle correlation 
measurements (Hanbury Brown/Twiss interferometry), 
which yield typically $v_f\simeq 0.5$ \cite{Heinz99}.
The thermophoretic force, being proportional to $R^2$, and consequently the 
flow would have been even stronger if we had assumed a larger radius $R$ for the 
hadronic clusters. Note also that the thermophoretic flow velocity is smaller for  
massive hadrons than for lighter ones. 

Here we assumed that the mixed phase lives for 10 - 20 fm/c for 
accelerating the hadrons over the distance $d$ from the center to the surface  
of the fireball. Such a long lifetime of the mixed phase would occur only
for a strong first order transition as predicted by simple hydrodynamical models
using an unrealistic equation of state (see e.g. \cite{Steffen01}). According to lattice 
results, showing at most a weak first order transition, this assumption is probably
a crude overestimation. However, even for small lifetimes of the mixed phase of 
1 - 2 fm/c, corresponding to $d\simeq 1$ fm, we obtain the same result for the final 
velocities if we take a moderate temperature gradient of 10 MeV/fm. This follows from
the fact that $W$ is proportional to $d$ times $dT/dr$. 
 
For our purpose the use of elementary kinetic theory, generalized here to relativistic 
plasmas, is justified since a more
elaborate treatment of thermophoresis based on the Boltzmann equation (Enskog-Chapman
method) \cite{Waldmann59} differs only by 20\% from the elementary kinetic result in 
the non-relativistic limit \cite{Rothermel01}. Similarly, the viscosity of the 
quark-gluon plasma
has been estimated from elementary kinetic arguments \cite{Thoma91} in good agreement
with a transport theoretical approach \cite{Baym90}.

We have seen that the thermophoretic force on hadrons in the mixed phase,
caused by a small to moderate temperature gradient, leads to a strong flow in addition 
to the hydrodynamical flow. This result is in contrast 
to present observations at SPS \cite{Jacak01}. Also the thermophoretic flow velocity
is larger for
lighter than for heavier hadrons, which contradicts experimental results showing a higher 
flow velocity for baryons than for mesons.
We take this as an indication for 
the absence of a mixed phase in relativistic heavy-ion collisions, investigated so far, 
i.e. either a continuous or second order phase transition or no phase transition at all,
also in aggreement with a rapid hadronization scenario \cite{Csernai95}. Anyway, the 
possible presence of a temperature gradient in a heavy-ion fireball 
could have other consequences which could be an interesting subject for future hydrodynamical
investigations, e.g. including dissipative effects such as heat conductivity.

\vspace*{1cm}

\centerline{\bf ACKNOWLEDGMENTS}
\vspace*{0.5cm}
The author would like to thank U. Heinz, P. Huovinen, and H. Rothermel for helpful discussions.


\begin{references}
\bibitem{Karsch01} F. Karsch, {\it hep-lat/0109017}.
\bibitem{Rischke96} D.H. Rischke and M. Gyulassy, Nucl. Phys. {\bf A608}, 479 (1996).
\bibitem{Csernai95} see e.g. L.P. Csernai and I.N. Mishustin, Phys. Rev. Lett. {\bf 74},
5005 (1995) and references therein. 
\bibitem{Bouchoule99} A. Bouchoule (Ed.), {\it Dusty Plasmas} (John Wiley, 1999).
\bibitem{Huovinen01} P. Huovinen, P.V. Ruuskanen, and S.S. R\"as\"anen, {\it nucl-th/0111052}.
\bibitem{Rothermel01} H. Rothermel, T. Hagl, G. Morfill, and H.M. Thomas,  
{\it physics/0110045}.
\bibitem{Huovinen02} P. Huovinen, private communication.
\bibitem{Ollitrault98} J.-Y. Ollitrault, Nucl. Phys. {\bf A638}, 195c (1998).
\bibitem{Reif65} F. Reif, {\it Fundamentals of Statistical and Thermal 
Physics} (McGraw-Hill, 1965).
\bibitem{Heinz99} U. Heinz, Nucl. Phys. {\bf A661}, 140c (1999). 
\bibitem{Steffen01} F.D. Steffen and M.H. Thoma, Phys. Lett. B {\bf 510}, 98 (2001).
\bibitem{Waldmann59} L. Waldmann, Z. Naturf. A {\bf 14}, 590 (1959).
\bibitem{Thoma91} M.H. Thoma, Phys. Lett. B {\bf 269}, 144 (1991).
\bibitem{Baym90} G. Baym, H. Monien, C.J. Pethick, and D.G. Ravenhall, Phys. Rev. Lett.
{\bf 64}, 1867 (1990).
\bibitem{Jacak01} J.M. Burward-Hoy and B.V. Jacak, {\it nucl-ex/0111005}.
 
\end{references}
\end{document}